\documentclass[12pt,preprint]{aastex}
\slugcomment{} 
\shorttitle{SDSS J1334$+$3315: A New Lensed Quasar}
\shortauthors{}
\begin{document}
\title{SDSS J133401.39+331534.3: A New Subarcsecond Gravitationally Lensed Quasar\altaffilmark{1}}
%
\author{
Cristian E. Rusu,\altaffilmark{2,3} 
Masamune Oguri,\altaffilmark{4,5} 
Naohisa Inada,\altaffilmark{6} 
Issha Kayo,\altaffilmark{7} 
Masanori Iye,\altaffilmark{2,3,8} 
Yutaka Hayano,\altaffilmark{9} 
Shin Oya,\altaffilmark{9} 
Masayuki Hattori,\altaffilmark{9} 
Yoshihiko Saito,\altaffilmark{9} 
Meguru Ito,\altaffilmark{9} 
Yosuke Minowa,\altaffilmark{9} 
Tae-Soo Pyo,\altaffilmark{9} 
Hiroshi Terada,\altaffilmark{9} 
Hideki Takami,\altaffilmark{9} 
and
Makoto Watanabe\altaffilmark{10}
}

\altaffiltext{1}{Based on data collected at Subaru Telescope, which is
operated by the National Astronomical Observatory of Japan. Use of
the UH2.2 m telescope for the observations is supported by NAOJ.} 
\altaffiltext{2}{Optical and Infrared Astronomy Division, National
Astronomical Observatory of Japan, 2-21-1,
Osawa, Mitaka, Tokyo 181-8588, Japan.}  
\altaffiltext{3}{Department of Astronomy, Graduate School of Science,
University of Tokyo 7-3-1, Hongo Bunkyo-ku, Tokyo 113-0033 , Japan} 
\altaffiltext{4}{Institute for the Physics and Mathematics of the
     Universe, The University of Tokyo, 5-1-5 Kashiwa-no-ha, Kashiwa, 
     Chiba 277-8568, Japan.}
\altaffiltext{5}{Division of Theoretical Astronomy, National
                Astronomical Observatory of Japan, 2-21-1, Osawa,
                Mitaka, Tokyo 181-8588, Japan.} 
\altaffiltext{6}{Department of Physics, Nara 
	      National College of Technology, Yamatokohriyama,
	      Nara 639-1080, Japan}    
\altaffiltext{7}{Department of Physics, Toho University, Funabashi, Chiba
	       274-8510, Japan.}
\altaffiltext{8}{Department of Astronomical Science, The Graduate University
                for Advanced Studies (SOKENDAI), National Astronomical Observatory
                of Japan, 2-21-1,Osawa, Mitaka, Tokyo 181-8588, Japan} 
\altaffiltext{9}{Subaru Telescope, National Astronomical Observatory of Japan,
	       650 North A'ohoku Place, Hilo, Hawaii 96720, USA} 
\altaffiltext{10}{Department of Cosmosciences, Hokkaido University,
	       Kita 10, Nishi 8, Kita-ku, Sapporo, Hokkaido 060-0810, Japan} 

\begin{abstract}
The quasar SDSS~J133401.39+331534.3 at $z=2.426$ is found to be a
two-image gravitationally lensed quasar with the image separation of
$0\farcs833$. The object is first identified as a lensed quasar candidate in 
the Sloan Digital Sky Survey Quasar Lens Search, and then confirmed as
a lensed system from follow-up observations at the Subaru and University of
Hawaii 2.2-meter telescopes. We estimate the redshift of the lensing galaxy
to be $0.557$ based on absorption lines in the quasar spectra as well as the
color of the galaxy. In particular, we observe the system 
with the Subaru Telescope AO188 adaptive optics with laser guide star, in 
order to derive accurate astrometry, which well demonstrates the usefulness
of the laser guide star adaptive optics imaging for studying strong
lens systems. Our mass modeling with improved astrometry implies that
a  nearby bright galaxy $\sim 4''$ apart from the lensing galaxy is likely to
affect the lens potential. 
\end{abstract}


\section{Introduction}\label{sec:intro}

Since the discovery of the first gravitationally lensed quasar
\citep{walsh79}, it has convincingly been
demonstrated that lensed quasars provide insights into various research
fields in astrophysics and cosmology, as well as being a unique tool for
studying the dark universe. For instance, one can directly measure the
structure and substructure of lensing objects, including the distribution of dark
matter, through detailed observations and analysis of lensed
systems \citep[e.g.,][]{treu04,rusin05,chiba05}. Magnification by
gravitational lensing greatly enhances our ability to study quasar
host galaxies at high-redshifts \citep[e.g.,][]{peng06}. Moreover, 
the statistics of lensed quasars, as well as measurements of time
delays between lensed images, provide independent constraints on
cosmological parameters \citep[e.g.,][]{chae02,oguri07,oguri08,suyu10}.

High resolution imaging such as that provided by Adaptive Optics (AO) is
crucial for accurately characterizing lensed quasar systems. This is not only
the case for the small-separation systems ($\sim1''$), where accurate astrometry (prone
to the effect of atmospheric seeing) is necessary for constraining mass
models of the lensing galaxies \citep[e.g.,][]{sluse08}, but also for the
larger separation systems, where adaptive optics imaging can identify lens
substructures and faint extended lensed features such as arcs \citep[e.g.,][]{mckean07}.   

The Sloan Digital Sky Survey Quasar Lens Search
\citep[SQLS;][]{oguri06,oguri08,inada08,inada10} 
has been one of the most successful strong lens surveys conducted to
date. SQLS relies on the large homogeneous sample of
spectroscopically-confirmed quasars from the Sloan Digital Sky Survey
\citep[SDSS;][]{york00}. By identifying lensed quasar candidates using
the color and morphological information from the SDSS imaging data,
SQLS has discovered more than 40 new lensed quasars so far. Together
with previously known lenses which were re-observed by the
SDSS, the SQLS sample consists of $\sim 60$ lensed quasars,
constituting roughly half of all lensed quasars discovered to date. 

However, a disadvantage of SQLS, like other ground-based optical
strong lens surveys, is its poor capability of identifying
subarcsecond (image separation $\theta<1''$) lensed systems. This is
because typical seeing sizes of ground-based optical imaging
observations are $\sim 1''$, which makes it difficult to resolve
multiple components of subarcsecond systems. Indeed, discoveries and 
confirmations of subarcsecond lensed quasars have been made mostly by
radio surveys \citep[e.g.,][]{browne03} or by using the Hubble Space
Telescope \citep[e.g.,][]{reimers02}, with few exceptions from
ground-based surveys \citep[e.g.,][]{castander06,blackburne08}. 

In this paper, we report the discovery of the first new subarcsecond (two-image)
lensed quasar discovered in the course of SQLS, SDSS~J133401.39+331534.3
(hereafter SDSS~J1334+3315). After briefly describing the lens
candidate selection from the SDSS data (\S\ref{sec:sdss}), we present
our imaging and spectroscopic follow-up observations (\S\ref{sec:obs}).
We further observe this system with the Subaru Telescope laser guide
star adaptive optics (LGS$+$AO188), for robust detection as well as
characterization of the lensing galaxy (\S\ref{sec:lgsao},
\S\ref{sec:photoz}). The accurate astrometry obtained with the LGS$+$AO188
 is then used for gravitational lens mass modeling (\S\ref{sec:mass}). We
summarize our results in \S\ref{sec:summary}. We assume the
concordance cosmology with $H_0=70 \ \mathrm{km}^{-1}\
\mathrm{s}^{-1}\ \mathrm{Mpc}^{-1}$, $\Omega_m=0.27$ and
$\Omega_\Lambda=0.73$ throughout this paper. 

\section{Candidate Selection from the Sloan Digital Sky 
Survey}\label{sec:sdss}

The SDSS \citep[SDSS-I and SDSS-II Legacy Survey;][]{york00} is a
combination of imaging  and spectroscopic surveys to map 10,000 square
degrees of the sky, centered at the North Galactic Cap.  
The survey uses a dedicated wide-field 2.5-meter telescope
\citep{gunn06} at the Apache Point Observatory in New Mexico, USA.  
Images taken in five broad-band filters
\citep[$ugriz$,][]{fukugita96,gunn98,doi10} are reduced with an automated
pipeline, achieving astrometric accuracy better than about
$0\farcs1$ \citep{pier03}, and photometric zeropoint accuracy better
than about 0.01 magnitude over the entire survey area, in the $g$, $r$, and $i$ bands 
\citep{hogg01,smith02,ivezic04,tucker06,padmanabhan08}. 
The SDSS also conducts spectroscopic observations with a multi-fiber
spectrograph covering  3800{\,\AA} to 9200{\,\AA}, with a resolution of
$R\sim1800$ \citep{blanton03}. The data are made publicly available 
\citep{stoughton02,abazajian03,abazajian04,abazajian05,abazajian09,
adelman06,adelman07,adelman08}. 

\begin{figure}
\epsscale{.5}
\plotone{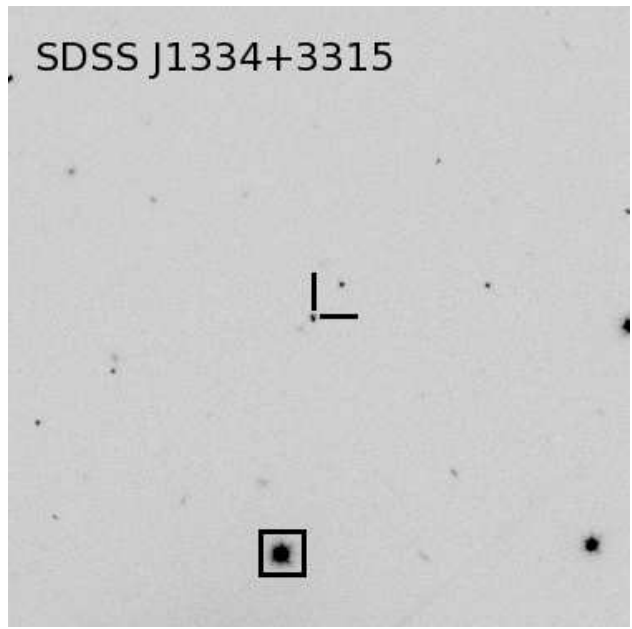}
\caption{Finding chart of SDSS~J1334+3315 from the SDSS $i$-band
image. The size is $2\farcm5\times 2\farcm5$. North is up and East
is left. The bright star indicated by an open square was used as the tip-tilt
star for the laser guide star adaptive optics imaging (see
\S\ref{sec:lgsao} for details).
\label{fig:fc_sdss1334}}
\end{figure}

The SQLS \citep{oguri06,oguri08,inada08,inada10} identifies lensed
quasar candidates among spectroscopic quasars in the SDSS
\citep{richards02}, using two selection algorithms. One is
the morphological selection, which selects quasars with
possible extended morphology as small-separation lensed candidates. 
The other is the color selection, which searches for nearby
stellar objects whose colors are similar to the spectroscopically
identified quasars. SDSS~J1334+3315 is a lensed quasar candidate
at $z=2.426$ selected by the morphological selection (see
Figure~\ref{fig:fc_sdss1334} for a finding chart). Although the
morphological selection is designed to be nearly complete at
$\theta>1''$, it can identify some subarcsecond  lensed systems as well,
as explicitly shown in simulations of lensed quasar images
\citep{oguri06}. The point spread function (PSF) magnitudes of the 
unresolved SDSS~J1334+3315 system, measured by the SDSS, are 
$u=19.32\pm0.03$, $g=18.76\pm0.01$, $r=18.66\pm0.02$, $i=18.80\pm0.02$, 
and $z=18.71\pm0.03$, where the Galactic extinction is not corrected.   

\section{Follow-up Spectroscopy and imaging}\label{sec:obs}

\subsection{Spectroscopy}\label{sec:spec}

We conducted spectroscopic follow-up observations of SDSS~J1334+3315
with the Faint Object Camera and Spectrograph \citep[FOCAS;][]{kashikawa02}
at the Subaru 8.2-meter telescope on January 31, 2009. We adopted the 300B
grism and the L600 filter to cover the wavelength from 3700{\,\AA} to 6000{\,\AA}.
We obtained data with an exposure of 600~s using a $1''$ slit, with
$2\times2$ on-chip binning. The resulting spatial resolution is
$0\farcs208$~pixel$^{-1}$, and the spectral resolution is $R\sim 400$.
The observations were necessary in order to extract individual spectra of the
two stellar components (the two quasar images), which are unresolved in the original
SDSS spectroscopy. The one-dimensional spectra of the stellar components were extracted 
using standard IRAF tasks. We were unable to extract the spectrum of
a lensing galaxy (see \S\ref{sec:image}), due to its faint nature and the subarcsecond
separation of the quasar components. Although the two stellar
components are resolved reasonably well in the two-dimensional
spectrum, we adopted a deblending procedure developed by
\citet{pindor06} in extracting one-dimensional spectra, in order to
minimize the cross-contamination between the two components.  

\begin{figure}
\epsscale{.8}
\plotone{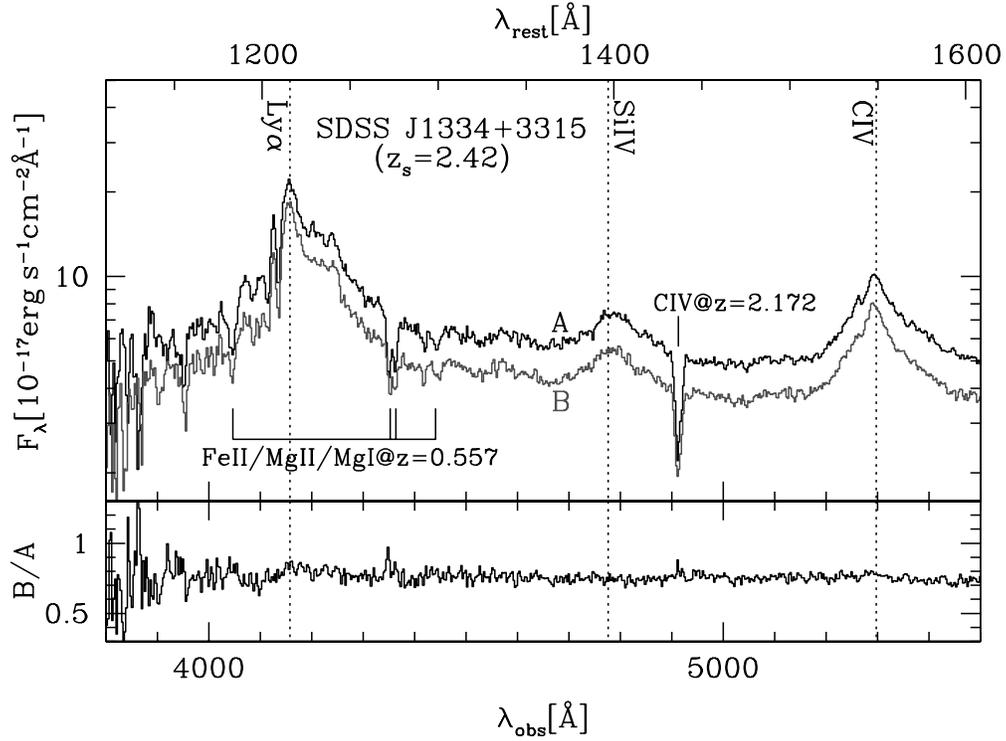}
\caption{Follow-up spectra of SDSS~J1334+3315 A and B taken with FOCAS
at the Subaru Telescope. Quasar emission lines redshifted to
$z_s=2.42$ are indicated by vertical dotted lines. Strong absorption 
systems at $z=0.557$ and $2.172$ seen in both component A and B are
marked by vertical solid bars. The lower panel plots the ratio of
the two spectra. 
\label{fig:spec1334}}
\end{figure}

The spectra shown in Figure~\ref{fig:spec1334} indicate that both 
stellar components have quasar broad emission lines (Ly$\alpha$, \ion{Si}{4},
\ion{C}{4}) at the same wavelengths, corresponding to a redshift of 2.426
(measured also by the initial SDSS spectroscopy). The emission lines
have the same overall shape, supporting the gravitational
lensing hypothesis. Several absorption lines are also observed in both
spectra, at the same wavelengths. We identify some of these as the
2599{\,\AA} \ion{Fe}{2} and the 2796{\,\AA}, 2803{\,\AA} \ion{Mg}{2}
doublet, from 
which we derive the redshift of the absorber to be $z=0.557$. The
absorption is seen in both components, with the absorption in
component A being slightly stronger than that in component B. We
postulate this absorber to be the lensing galaxy, as the redshift is
consistent with the photometric redshift estimate of the lensing
galaxy, as we will discuss later (\S\ref{sec:photoz}). In addition, we
identify a strong \ion{C}{4} absorber at $z=2.172$ in both stellar
components.   

The ratio of the fluxes of the two stellar components is quite
constant along the plotted range (B/A = 0.7 -- 0.8). There is
therefore no sign of differential dust reddening which, should it
occur on the line of sight to one of the quasars, would affect the
flux ratio towards shorter wavelengths. Because the flux ratio is
quite constant both across the continuum and the broad line regions,
there is also no evidence of microlensing effects. Microlensing due to
stars in the lensing galaxy would effect the continuum emission more
strongly than the broad line emission, since the physical size of the
region associated with the former is much smaller than the one associated
with the latter \citep[e.g.,][]{sluse08}. The change in flux
ratio below $\sim 4000${\,\AA} in the observer's frame is thought to be
due to errors arising from the low detector response in that region. 

\subsection{Optical and Near-Infrared Imaging}\label{sec:image}

Optical ($I$ and $z$) follow-up images were taken with the Tektronix 
2048$\times$2048 CCD camera (Tek2k) at the University of Hawaii
2.2-meter (UH88) telescope on 2009 April 15. The pixel scale is
$0\farcs219$~pixel$^{-1}$. The total exposure time was 800~s for each 
filter. The seeing was $\sim 1''-1\farcs2$. The images were reduced
using standard IRAF\footnote{IRAF is distributed by the National 
Optical Astronomy Observatories, which are operated by the
Association of Universities for Research in Astronomy, Inc., under
cooperative agreement with the National Science Foundation.} tasks.
The photometric zero-points were derived by comparing magnitudes of
nearby stars located in the same frame to their SDSS magnitudes,
 with the resulting zero-point accuracy of $\sim 0.05$~mag. For the
 $I$ band, the magnitude conversion has been performed using the
 equations of Lupton 2005.\footnote{\protect\url{http://www.sdss.org/dr5/algorithms/sdssUBVRITransform.html}}

In addition, we obtained near-infrared ($JHK_s$) images with
the Multi-Object InfraRed Camera and Spectrograph 
\citep[MOIRCS;][]{ichikawa06,suzuki08} at the Subaru 8.2-meter
telescope \citep{iye04} on 2010 April 2, under an excellent seeing size of $\sim
0\farcs4$. The pixel scale is $0\farcs117$~pixel$^{-1}$. The total
exposure time was 600~s for $J$ and $K_S$ bands, and 480~s for $H$
band. We use the MCSRED software package (I. Tanaka et al.,
in preparation) for the data reduction. We derived the magnitude
zero-points using the standard star FS33 \citep{leggett06}, leading to
a zero-point accuracy estimated at $\sim$ 0.05~mag. 

\begin{figure}
\epsscale{.6}
\plotone{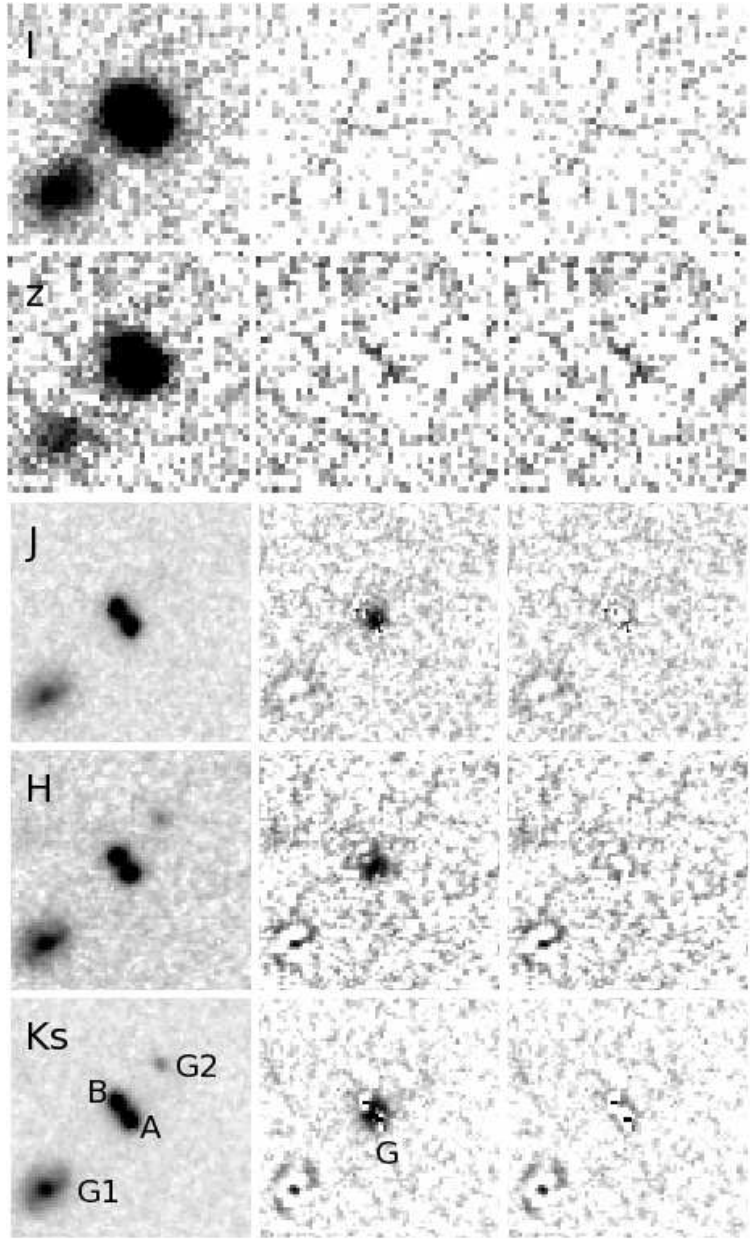}
\caption{Follow-up images taken with Tek2k at the UH88 Telescope 
and MOIRCS at the Subaru Telescope. The size of each panel is
$9''\times9''$. North is up and East is left.  From top to bottom
panels, we show $I$- and $z$-band images from Tek2k and $J$-, $H$-,
and $K_s$-band images from MOIRCS. Left, middle, and right panels
display original images, images after subtracting all but the
lensing galaxy (component G), and images after subtracting all
components, respectively. 
\label{fig:image_IzJHK}}
\end{figure}

Results of the follow-up imaging are summarized in
Figure~\ref{fig:image_IzJHK}. In addition to the candidate lensed quasar
system, there are two red galaxies (G1 and G2) within $5''$. In order
to check the presence of the lensing galaxy in between the stellar
components, as well as to extract relative astrometry and photometry
from the data, we analyze the images using the public software GALFIT
\citep{peng02}. We fit all components simultaneously with nearby stars
as PSF templates and galaxies modeled by the S\'{e}rsic profile. 
For the bright galaxy G1, the preliminary GALFIT fitting produces the
best-fit S\'{e}rsic index of $n=0.87$, very close to the 
canonical value $n=1$ for disk-like galaxies. The faint galaxy G2 appears to
be quite red, and is also not detected in the optical images. Due to its 
faintness, we cannot extract an accurate S\'{e}rsic index.
   
Table~\ref{tab:image} provides a summary of the GALFIT modeling
results. The separation between the stellar components (the two quasar
images), derived from the $K_s$-band image, is $0\farcs83\pm0\farcs01$. 
We detect an extended object (galaxy G) in between the two
stellar components (the two quasar images), which can be interpreted
as a lensing galaxy. We find that the lensing galaxy is hardly detected in the
UH88 optical images, while it is significantly detected in the Subaru/MOIRCS
near-infrared images (see also Figure~\ref{fig:image_IzJHK}). The
difference can be attributed to the much better angular
resolution and the deeper depth of the MOIRCS images. The position of
the lensing galaxy G is difficult to determine accurately, as the best fit varies
between different bands. However, by adopting the position estimated in the
$K_s$-band, where G appears brightest, we
determine that the center of the lensing galaxy is closer to the
brighter stellar component, a fact often seen among two-image  
lensed quasars \citep[e.g.,][]{kayo10}. We are unable to determine a robust
estimate of the S\'{e}rsic index for G, due to the galaxy being blended with
the quasar components. In order to extract magnitudes for G,  we adopt the
canonical value for elliptical galaxies, $n=4$. This is justified by
the fact that most galaxies acting as lenses are ellipticals, as
spirals only contribute to about 10\% of the probability for multiple
imaging \citep[e.g.,][]{fukugita91}. 

\section{Laser Guide Star Adaptive Optics Imaging}\label{sec:lgsao}

While the spectroscopic and imaging follow-up observations presented
in the previous section indicate that SDSS~J1334+3315 is a true
lensed quasar system, its small angular separation makes it difficult
to study the system in detail solely from these observations. One way
to overcome this problem in the near-infrared is to obtain high-resolution  
adaptive optics images, as already done for several other quasar lens
systems \citep[e.g.,][]{mckean07,auger08,sluse08,lagattuta10}. 

We obtained high-resolution near-infrared images for this quasar system
using the Infrared Camera and Spectrograph \citep[IRCS;][]{kobayashi00} at the 
Subaru telescope, along with the Laser Guide Star Adaptive Optics system
\citep[LGS$+$AO188;][]{hayano08,hayano10,minowa10}. AO188
uses a curvature sensor with 188 control elements and a 188 element bimorph
deformable mirror (BIM188), and operates at the Nasmyth focus of the Subaru
Telescope. IRCS was used in the low resolution mode with a pixel scale
$0\farcs0527$, producing a $54''$ field of view. The observations were
performed in an engineering run on 2011 February 18, with excellent
seeing $\sim 0\farcs3 - 0\farcs5$. We used an $R \sim13.4$~mag star
located $57''$ from the target as a tip-tilt guide star (see
Figure~\ref{fig:fc_sdss1334}). We obtained a set of 60~s 
individual exposures with a total exposure time of 10~min for $J$-band,
4~min for $H$-band, and 12~min in $K'$-band. The measured Strehl
ratios were $\sim 8\%$ in $J$-band, $\sim 12\%$ in
$H$-band, and $\sim 12\%$ in $K'$-band. The full width at half
maximum (FWHM) of the PSF was $\sim 0\farcs10 - 0\farcs15$ in all
three bands. Therefore, although the PSF is not quite diffraction limited ($\sim 0.06''$
for Subaru Telescope in the $K'$-band), it is better than what would be obtained
with the Hubble Telescope at diffraction limit, in the equivalent band ($\sim 0.22''$).

The standard star P272-D \citep{leggett06} was observed in the same
night and at the same airmass with the target, without adaptive
optics.  Both the target and standard star frames were reduced with
the IRCS IMGRED package \citep{minowa05}. The
zero-points which we estimate from aperture photometry are $J =
24.39\pm 0.01$~mag/s, $H = 24.53\pm 0.02$~mag/s, and $K' = 23.78\pm
0.02$~mag/s, respectively, where the quoted errors refer only to the
differences between aperture photometry of the standard star in
different frames. 
 
We initially tried to build the PSF on an isolated star $\sim 10''$
from the target, in the direction opposite to the tip-tilt star.
However, this left significant residuals when
subtracting the quasar components, most likely due to
anisoplanatism introduced by the  adaptive optics. We therefore built the
PSF by fitting a two-component model, consisting of the Moffat and Gaussian
analytical profiles, to the relatively isolated fainter quasar component.
Our GALFIT modeling suggests a best-fit Moffat
profile FWHM of $\sim 6 - 8$~pixels ($\sim 0\farcs3 - 0\farcs4$),
which is indicative of the seeing disk, and a Gaussian profile FWHM of
$\sim 2 - 3$~pixels ($\sim 0\farcs10 - 0\farcs15$), corresponding to
the PSF core. This method significantly improves fitting and reduces
the residuals after subtracting best-fit models. 

\begin{figure}
\epsscale{.7}
\plotone{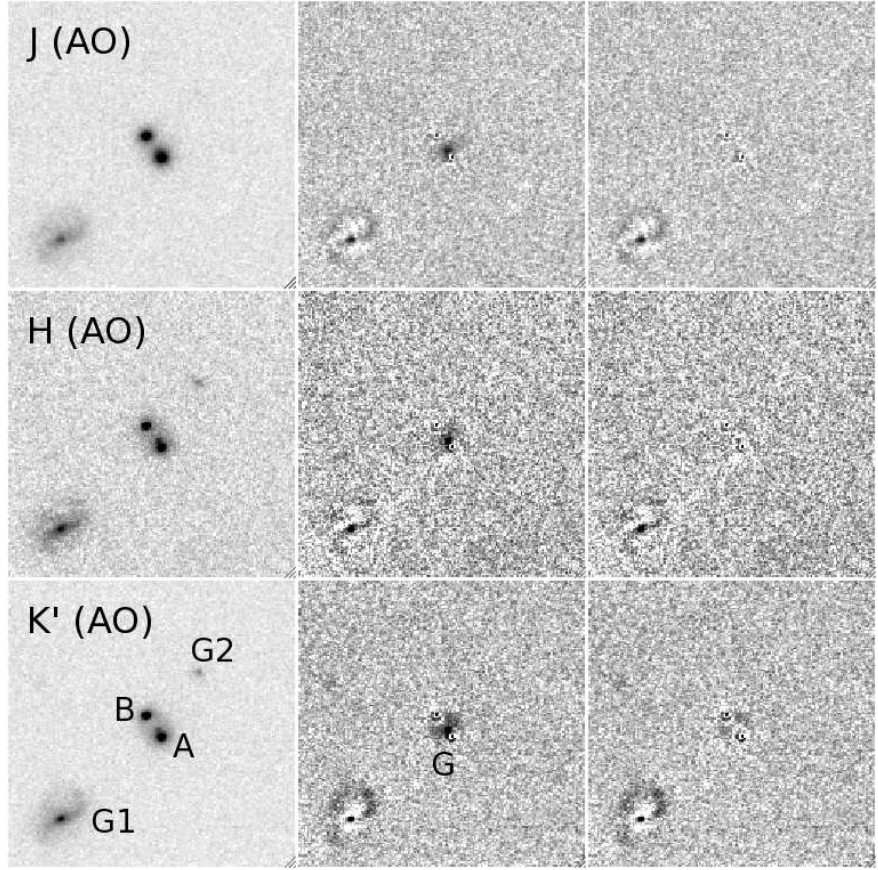}
\caption{Subaru LGS$+$AO188 images of SDSS~J1334+3315 in $J$-, $H$-, and
 $K'$-bands. The size of each panel is $9''\times9''$. North is up
 and East is left. As in Figure~\ref{fig:image_IzJHK}, from left to
 right panels we show original images, images after subtracting all
 but the lensing galaxy (component G), and images after subtracting
 all components, respectively. 
\label{fig:image_ao}}
\end{figure}

We show the adaptive optics images in Figure~\ref{fig:image_ao}, and
summarize the results we obtained in Table~\ref{tab:lgsao}. 
Thanks to the much improved angular resolution, the lensing galaxy is
clearly visible even before subtracting the stellar components. The
spiral arm of the nearby galaxy G1 is also clearly seen. The relative
astrometry derived from the LGS$+$AO188 image is broadly consistent with the
MOIRCS result shown in Table~\ref{tab:image}, but the attached errors are now
much smaller. There are small discrepancies in the relative positions of the quasar
images, larger than the error bars attached to the MOIRCS measurements, which 
must be due to an inaccurate original measurement of the position of image A (made
difficult by the close proximity of the lens galaxy). The LGS$+$AO188 measurement
corrects the MOIRCS-determined values, because the image A and the lens are
more clearly separated. In addition, small (sub-pixel) discrepancies between the MOIRCS-
and LGS$+$AO188-determined positions of galaxies G1 and G2 should be due to
better detected morphological features in the LGS$+$AO188 imaging. The quasar
image separation is found to be $\theta=0\farcs833\pm0.002$. We also confirm that 
the center of the lensing galaxy is closer to the brighter quasar component.  In addition,
the high-resolution images enable us to make a more reliable measurement of
the shape of each galaxy, such as the half-light radius and the
ellipticity. Although the goodness-of-fit estimate $\chi^2$ is almost insensitive to the
choice of fiducial S\'{e}rsic indices 1 (exponential profile) or 4 (de Vaucouleurs profile),
less residuals appear in the $K'$-band after subtracting the lens galaxy modeled by the
S\'{e}rsic index 4. This suggests an elliptical galaxy, in accordance with our original assumption.

We must note that the photometry (especially that of the
lensing galaxy) differs significantly from the MOIRCS result. This is
presumably because of the PSF and magnitude zero-point uncertainties,
which are hard to characterize for AO images. On the other hand, relative photometry
of the lens - quasar images should be more accurate in the case of the AO observations,
due to the higher resolution. Also, objects (especially the galaxies) appear systematically
fainter in the AO images, which may be due to uncorrected light 
scattered at large distances. While we leave for
future work more comprehensive and sophisticated analysis of the PSF
and photometric calibration, we confirmed that our astrometric results
are rather robust against these uncertainties. 

\section{Photometric redshifts}\label{sec:photoz}

In order to check the validity of our assumption that the strong
absorption at $z=0.557$ seen in the quasar spectra is caused by the
lensing galaxy G, we estimate a photometric redshift for this galaxy 
based on the magnitudes in the five observed bands ($I$, $z$, $J$,
$H$, $K_s$/$K'$). We also estimate photometric redshifts for the two
nearby galaxy companions. We employ the publicly available HyperZ
\citep{bolzonella03} and EAzY \citep{brammer08} algorithms, both of
which produce redshift estimates based on template-fitting
methods. For both algorithms, we use as templates the observed mean
spectral energy distributions (SEDs) of local galaxies from
\citet[][hereafter CWW]{coleman80}, extrapolated into ultraviolet and
near-infrared with the evolutionary models of \citet{bruzual93}. In
the case of EAzY, we use a Bayesian prior on the apparent magnitudes,
which is implemented in the code. The prior is added to help prevent
the redshift probability distributions from having multiple peaks, as
template colors can be degenerate with redshift.  
  
Since the best magnitude estimates for the lensing galaxy are
significantly different between the MOIRCS $JHK_s$- and the
LGS$+$AO188 IRCS $JHK'$-bands, we use both of these estimates separately. 
On the other hand, for the two nearby galaxies, where the aperture
magnitudes are fairly similar, we use the values obtained with LGS$+$AO188. 
As photometric redshift estimates require absolute magnitudes, we attach error
bars larger than the ones quoted in Tables \ref{tab:image} and \ref{tab:lgsao}, 
to allow for uncertainties in the magnitude zero-points.
 
\begin{figure}
\epsscale{0.9}
\plotone{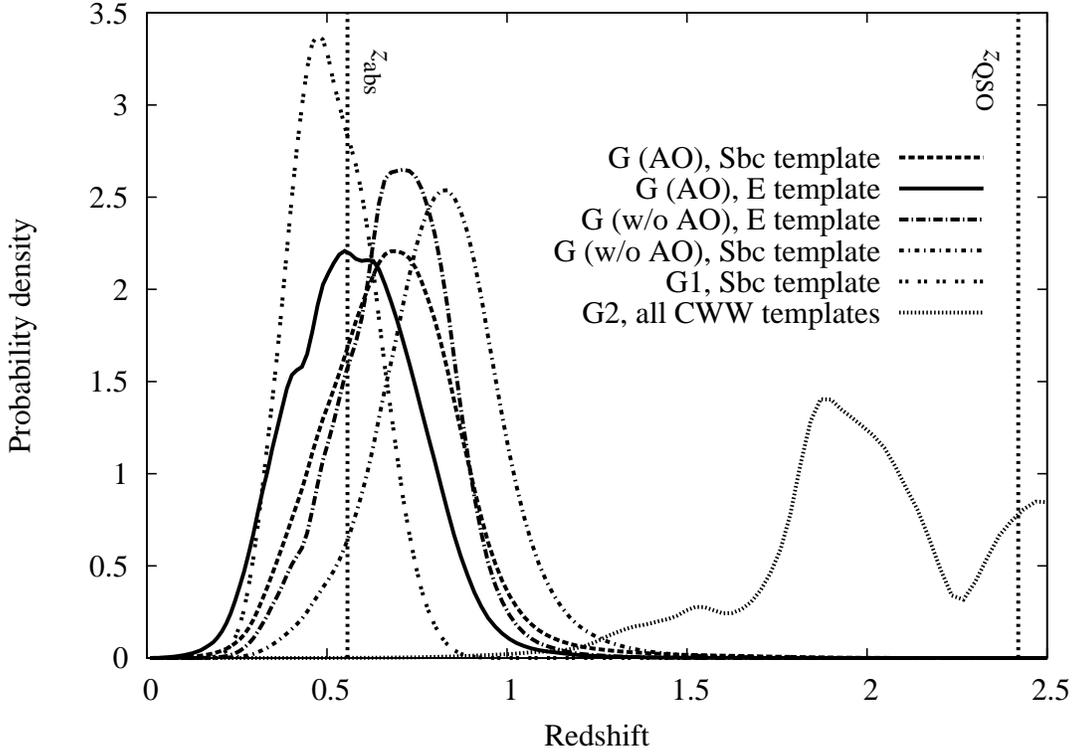}
\caption{Photometric redshift probability distributions, normalized to unit area,
for the three galaxies G, G1 and G2, calculated using the magnitudes in Tables \ref{tab:image}
and \ref{tab:lgsao}. In the case of G, we consider separately the IRCS $JHK'$
(with Adaptive Optics) and MOIRCS $JHK_s$ (w/o AO) magnitudes. The probability distributions are
 calculated with EAzY, for the most consistent template estimates (see
\S\ref{sec:photoz} for details). The redshift of the
 quasar components, as well as that of the absorption line system
 measured in both quasars at $z=0.557$, are also marked with vertical lines.
\label{fig:photoz_ao}}
\end{figure}

We conducted initial photometric redshift fits with HyperZ, and checked
the results of the best-fitted spectral templates with EAzY. The
resulting photometric redshift probability distributions computed with EAzY are
shown in Figure~\ref{fig:photoz_ao}, and summarized in Table~\ref{tab:photoz}. For the lensing galaxy, the
best-fitted templates obtained with HyperZ are E (elliptical) and Sbc
(spiral). The results are consistent with the probability curves
calculated by EAzY. Although we cannot conclusively discriminate between the two
templates based on the morphological information obtained even with
LGS$+$AO188, both templates yield redshift probability distributions
consistent with the redshift estimated from the absorption lines,
within $1\sigma$ confidence interval when using the LGS$+$AO188-estimated
magnitudes, and at $1-2\sigma$ confidence interval when using
MOIRCS-estimated magnitudes. We therefore conclude that the lensing galaxy
is likely to be at $z=0.557$, and is certainly at a redshift lower than that of the
quasar components, consistent with the gravitational lensing hypothesis.
   
For the bright nearby galaxy G1, the probability curve for the Sbc
template (matching the disk-like morphology) is in good agreement
with $z=0.557$, making it very likely that this galaxy is located at
the same redshift as the lensing galaxy. It is also in good agreement
with the previous redshift estimate of $0.50 \pm 0.07$, from the SDSS
database. The G2 galaxy has a very poorly estimated redshift, due to 
its faintness, and also appears to be located at a redshift much
larger than G and G1. 

\section{Mass Modeling}\label{sec:mass}

We conduct gravitational lensing mass modeling to check whether the
observed image configuration can be reproduced with reasonable mass
models, as a final check of the gravitational lensing hypothesis for
this system.  We use the public software {\it glafic} \citep{oguri10}
to solve the lens equation and to find best-fit mass models. We use 7
observational constraints for mass modeling: the (two-dimensional) positions 
of the lensed quasar images and the lensing galaxy, as well as the flux ratio 
between the quasar images. The positional constraints are taken from the accurate
astrometric results derived from Subaru LGS$+$AO188 imaging
(Table~\ref{tab:lgsao}). We use a flux ratio constraint of
$B/A=0.75\pm0.05$, consistent with most  follow-up photometric results
as well as the spectroscopic flux ratio over a wide wavelength
range. The lens redshift is assumed to be $z=0.557$. Given the small
number of observational constraints available, we are forced to
consider the simplest mass models: the Singular Isothermal Ellipsoid
(SIE) and the Singular Isothermal Sphere with an external shear
(SIS+$\gamma$). Both mass models have 7 parameters (the position of 
the lens and its velocity dispersion $\sigma$ or Einstein radius $R_{Ein}$, 
the ellipticity $e$ or shear $\gamma$ and their associated position angles 
$\theta_e$ or $\theta_\gamma$, and the position of the source), which indicates
that there are no degree of freedom, as is common in the mass
modeling of two-image quasar lenses.  

We find that both models can fit the lensed system perfectly, 
signifying that the choice of the models is reasonable.
Table~\ref{tab:model} summarizes the best-fit values for both models.
We find that the best-fit ellipticity for the SIE model, $e=0.235$, is
significantly smaller than the observed ellipticity of the lensing
galaxy, $e \sim 0.62$, but the position angles agree quite well with each
other. Such large difference of the ellipticity and the agreement of
the position angle has often been seen in previous analyses of strong
lens systems \citep[e.g.,][]{keeton98}. The position angle of the
best-fit SIS+$\gamma$ model is almost the same as that of the SIE model. In
fact the angle corresponds to the position of the nearby galaxy G1, 
suggesting that the quadrupole of the lens potential may be dominated
by perturbation due to G1. For both models, the total magnification
and time delay are predicted to be $\mu_{\rm tot}\sim 6.0$ and $\Delta
t\sim 10 - 11$~days, respectively. 

Here we discuss whether galaxy G1 can have a significant impact on the
mass modeling. Based on the analysis in Section~\ref{sec:photoz}, we
assume that the redshift of galaxy G1 is the same as that of the lensing
galaxy. The follow-up photometry indicates that G1 is at least $\sim
1.5$~mag brighter than the lensing galaxy. Adopting the Faber-Jackson
relation, we infer that the Einstein radius of G1 is at least twice as
large as that of the lensing galaxy, i.e., $R_{\rm Ein}\sim 0\farcs8$. 
Given the distance between G and G1 of $d \sim 4\farcs1$, we estimate the
external shear produced by G1 as $\gamma\sim R_{\rm Ein}/2d\sim 0.1$, which
is very close to the best-fit external shear of $\gamma\sim 0.096$ for the
SIS+$\gamma$ model, suggesting that G1 can have a significant impact
on this lensed quasar system. A caveat is that G1 appears to be a disk-dominated
spiral galaxy (see Figure~\ref{fig:image_ao} and Table~\ref{tab:lgsao}), 
and hence the use of the Faber-Jackson relation may not be appropriate. 

\begin{figure}
\epsscale{.8}
\plotone{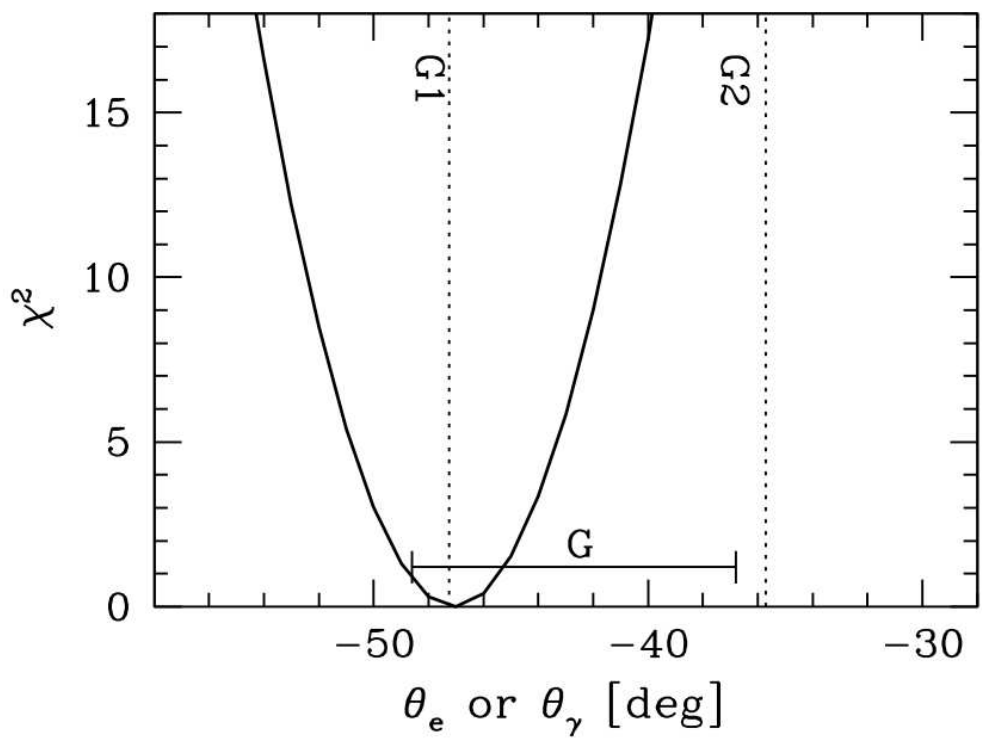}
\caption{The best-fit $\chi^2$ as a function of the position angle
measured East of North. Since the results are very similar for SIE
and SIS+$\gamma$ models, the position angle can be interpreted as
either $\theta_e$ of SIE or $\theta_\gamma$ of SIS+$\gamma$. Note
that the other model parameters are marginalized over. The horizontal
line indicates the $1\sigma$ range of the measured position angle
of the lensing galaxy G. The vertical dotted lines show directions of
the external shear corresponding to the location of nearby galaxies
G1 and G2. 
\label{fig:model_pa}}
\end{figure}

To further explore the origin of the quadrupole, we compute the
goodness-of-fit estimate $\chi^2$ as a function of the position
angle. Specifically we compute $\chi^2$ for each fixed value of the
position angle, with the other model parameters varied to achieve a
minimum $\chi^2$. We find that the results are very similar between
the position angle of the ellipticity $\theta_e$, for the SIE model,
and the position angle of the external shear $\theta_\gamma$, for the
SIS+$\gamma$ model, as a consequence of the well-known
shear-ellipticity degeneracy. The result shown in
Figure~\ref{fig:model_pa} indicates that the position angle inferred
from mass modeling is perfectly consistent with an external shear from galaxy
G1, as well as consistent with the position angle of the elliptical lensing galaxy
G at $1\sigma$ level. On the other hand, the location of the nearby
galaxy G2 corresponds to an external shear with a large $\chi^2$,
which implies that G2 is not affecting the lens potential. While we
cannot decompose contributions from G and G1 at this point, detailed
observations of a lensed host galaxy in deeper adaptive optics imaging
observations may help clarify the origin of the quadrupole.  

\section{Summary}\label{sec:summary}

We have reported the discovery of the subarcsecond ($\theta=0\farcs833$) 
gravitationally lensed quasar SDSS~J1334+3315. The system has been
identified as a lensed candidate from the SDSS data, using the standard
SQLS candidate selection algorithm, despite the image separation being
smaller than the criterion for which the morphological selection is complete.
Our follow-up observations at the Subaru and UH88 telescopes have
confirmed the system to be a real strong lensing event. It consists of
two images of a quasar at $z=2.426$, lensed by a foreground galaxy. 
From the colors as well as absorption lines in the quasar spectra, we
infer the redshift of the lensing galaxy to be $z=0.557$. In
particular, we have obtained high-resolution near-infrared images of
this lensed quasar system using the Subaru LGS$+$AO188 system, with a better resolution in
the $K'-$band than would be provided by the Hubble space telescope in the
equivalent band. The images clearly reveal the presence of the lensing galaxy
in between the quasar images, and enable us to derive very accurate relative
astrometry, as well as shapes of galaxies. In particular, the higher resolution made
possible to identify errors in the non-adaptive optics astrometry, caused by the
lens galaxy and one of the quasar images being blended, and thus made possible
a more accurate mass modeling. We have found that
the image configuration can well be reproduced with the standard mass models. 
We have pointed out that the nearby galaxy G1, which is located $<5''$ from the
lensing galaxy and is likely associated physically with the lensing
galaxy, may affect the lens potential. This lensed system has the
smallest image separation among the newly discovered SQLS lenses.  

This work represents the first observation and analysis of any
gravitationally lensed quasar system with the Subaru LGS$+$AO188. Our
results demonstrate the usefulness of laser guide star adaptive optics imaging observations, in
particular that of the Subaru Telescope LGS$+$AO188 imaging, which has a much
improved sky coverage, for the study of strong lens systems. As
explicitly shown, the high-resolution imaging enabled by
the adaptive optics system is crucial for accurate and robust measurements of
the position and shape of the lensing galaxy, particularly for
small-separation lensed quasars as presented in the paper. Indeed we are
conducting an imaging survey of SQLS lenses with Subaru LGS$+$AO188, aiming
to derive accurate astrometry and to detect lensed quasar host
galaxies, which will further increase the value of the SQLS lensed quasar sample. 

\acknowledgments

CER is sponsored by the Japanese Monbukagakusho scholarship.
I.K. acknowledges the support of the JSPS Research Fellowship.
This work was supported in part by the FIRST program "Subaru
Measurements of Images and Redshifts (SuMIRe)" and World Premier
International Research Center Initiative (WPI Initiative), MEXT,
Japan.  
The Subaru LGS$+$AO188 project has been supported by the Grant-in-Aid for
Specially Promoted Research 14002009 from the Japan Ministry of Education, 
Culture, Sports, Science, and Technology and by the Grant-in-Aid for 
Scientific research (S)19104004 from Japan Society for the Promotion of Science.
The authors recognize and acknowledge the very significant cultural
role and reverence that the summit of Mauna Kea has always had within
the indigenous Hawaiian community. We are most fortunate to have the
opportunity to conduct observations from this superb mountain. 

Funding for the SDSS and SDSS-II has been provided by the Alfred
P. Sloan Foundation, the Participating Institutions, the National
Science Foundation, the U.S. Department of Energy, the National
Aeronautics and Space Administration, the Japanese Monbukagakusho, the
Max Planck Society, and the Higher Education  Funding Council for
England. The SDSS Web Site is http://www.sdss.org/. 

The SDSS is managed by the Astrophysical Research Consortium for the
Participating Institutions. The Participating Institutions are the
American Museum of Natural History, Astrophysical Institute Potsdam,
University of Basel, Cambridge University, Case Western Reserve
University, University of Chicago, Drexel University, Fermilab, the
Institute for Advanced Study, the Japan Participation Group, Johns
Hopkins University, the Joint Institute for Nuclear Astrophysics, the
Kavli Institute for Particle Astrophysics and Cosmology, the Korean
Scientist Group, the Chinese Academy of Sciences (LAMOST), Los Alamos
National Laboratory, the Max-Planck-Institute for Astronomy (MPIA),
the Max-Planck-Institute for Astrophysics (MPA), New Mexico State
University, Ohio State University, University of Pittsburgh,
University of Portsmouth, Princeton University, the United States
Naval Observatory, and the University of Washington.

\clearpage

\begin{deluxetable}{crrccccc}
\tablewidth{0pt}
\rotate
\tabletypesize{\footnotesize}
\tablecaption{Result of imaging observations with UH88/Tek2k and Subaru/MOIRCS}
\tablewidth{0pt}
\tablehead{\colhead{Name} & 
\colhead{{$\Delta$}{\rm X} [arcsec]} &
\colhead{{$\Delta$}{\rm Y} [arcsec]} &
\colhead{$I$} & \colhead{$z$} & 
\colhead{$J$} & \colhead{$H$} & \colhead{$K_s$} }
\startdata
A & $0.000\pm0.003$  & $0.000\pm0.003$ & $19.16\pm0.03$ & $19.18\pm0.03$ & $18.57\pm0.02$ & $18.10\pm0.02$ & $17.41\pm0.01$ \\
B &$-0.502\pm0.003$  & $0.673\pm0.003$ & $19.50\pm0.02$ & $19.75\pm0.03$ & $18.99\pm0.02$ & $18.40\pm0.01$ & $17.72\pm0.01$ \\
G &$-0.159\pm0.015$  & $0.235\pm0.020$ & $21.50\pm0.37$ & $21.92\pm1.69$ & $20.29\pm0.11$ & $19.30\pm0.07$ & $18.53\pm0.06$ \\
G1&$-3.160\pm0.004$  &$-2.587\pm0.004$ & $19.95\pm0.04$ & $20.33\pm0.07$ & $18.68\pm0.03$ & $17.70\pm0.02$ & $17.06\pm0.02$ \\
G2& $1.160\pm0.011$  & $2.073\pm0.013$ & \nodata        & \nodata        & $22.18\pm0.18$ & $20.42\pm0.08$ & $19.69\pm0.07$ \\
\enddata
\tablecomments{The positive directions of {$\Delta$}{\rm X} and
{$\Delta$}{\rm Y} are defined by West and North, respectively. The astrometry is
measured in the $K_s$-band image. The quoted errors include
statistical errors only. The magnitude measurements for galaxies G1 and G2 are
 photometric aperture magnitudes, whereas those of components A, B,
 and G are model estimates fitted with GALFIT. The quoted errors
 include statistical errors only, and exclude uncertainties inherent
 in the PSF estimates or the zero-points. Magnitudes are given in the
 Vega system, without the Galactic extinction correction, except
 for $z$-band magnitudes which are in the AB system}   
\label{tab:image}
\end{deluxetable}

\begin{deluxetable}{crrccccccc}
\tablewidth{0pt}
\rotate
\tabletypesize{\footnotesize}
\tablecaption{Result of Subaru LGS$+$AO188 imaging}
\tablewidth{0pt}
\tablehead{\colhead{Name} & 
\colhead{{$\Delta$}{\rm X} [arcsec]} &
\colhead{{$\Delta$}{\rm Y} [arcsec]} &
\colhead{$n$} & \colhead{$R_e$ [arcsec]} & $e$ & \colhead{$\theta_e$ [deg]} &
\colhead{$J$} & \colhead{$H$} & \colhead{$K'$} }
\startdata
A  & $0.000\pm0.001$ & $0.000\pm0.001$ &      \nodata &       \nodata &       \nodata &   \nodata     & $18.71\pm0.01$ & $18.27\pm0.01$ & $17.53\pm0.01$ \\
B  & $-0.484\pm0.001$& $0.677\pm0.001$ &      \nodata &       \nodata &       \nodata &   \nodata     & $19.03\pm0.01$ & $18.54\pm0.02$ & $17.83\pm0.01$ \\
G  & $-0.134\pm0.006$& $0.235\pm0.007$ & $\equiv4$    & $0.20\pm0.03$ & $0.62\pm0.08$ & $-42.7\pm5.9$ & $21.01\pm0.06$\tablenotemark{a} & $20.06\pm0.05$\tablenotemark{a} & $19.36\pm0.05$ \\
G1 &$-3.178\pm0.004$ &$-2.578\pm0.002$ & $1.68\pm0.06$& $0.50\pm0.01$ & $0.53\pm0.01$ & $-58.2\pm0.7$ & $18.68\pm0.04$ & $17.90\pm0.03$ & $17.28\pm0.03$ \\
G2 & $1.159\pm0.005$ & $2.033\pm0.004$ & $2.46\pm1.09$& $0.08\pm0.01$ & $0.61\pm0.11$ &  $38.3\pm7.0$ & $22.69\pm0.31$ & $20.52\pm0.13$ & $19.94\pm0.10$ \\
\enddata
\tablecomments{The positive directions of {$\Delta$}{\rm X} and
 {$\Delta$}{\rm Y} are defined by West and North, respectively, and
 the astrometry is estimated from the $K'$-band image. Shapes of the
 galaxies (the S\'{e}rsic index $n$, the half-light radius $R_e$, the
 ellipticity $e$, and the position angle $\theta_e$ measured East of
 North) are all measured in the $K'$-band image. The S\'{e}rsic index
 of galaxy G is considered to be the fiducial value of 4, as explained
 in the text. The magnitude measurements for galaxies G1 and G2 are
 photometric aperture magnitudes, whereas those of components A, B,
 and G are model estimates fitted with GALFIT. The quoted errors
 include statistical errors only, and exclude uncertainties inherent
 in the PSF estimates or the zero-points. Magnitudes are given in the
 Vega system, without the Galactic extinction correction. }    
\label{tab:lgsao}
\tablenotetext{a}{The quoted values of the lens galaxy magnitudes in
 the $J$- and $H$- bands correspond to the best-fit models (S\'{e}rsic index 1
 rather than 4). When constraining an index of 4, the values become 
 $\sim 0.1$ mag lower.)}
\end{deluxetable}

\begin{deluxetable}{lcccccc}
\tablewidth{0pt}
\tabletypesize{\footnotesize}
\tablecaption{Best-fit photometric redshifts}
\tablewidth{0pt}
\tablehead{\colhead{Object \& template} & 
\colhead{Best-fit $z$} &
\colhead{$1\sigma$ limits} &
\colhead{$2\sigma$ limits} }
\startdata
G (AO), Sbc template & $0.694$  & $0.494 - 0.876$ & $0.329 - 1.174$ \\
G (AO), E template & $0.586$  & $0.397 - 0.768$ & $0.255 - 0.959$ \\
G (w/o AO), E template & $0.694$  & $0.530 - 0.850$ & $0.368 - 1.004$ \\
G (w/o AO), Sbc template & $0.818$  & $0.644 - 0.986$ & $0.456 - 1.202$ \\
G1, Sbc template & $0.513$  & $0.390 - 0.642$ & $0.298 - 0.757$ \\
G2, all CWW templates & $2.124$  & $1.774 - 2.579$ & $1.311 - 2.884$ \\
\enddata
\label{tab:photoz}
\end{deluxetable}

\begin{deluxetable}{ccccccc}
\tablewidth{0pt}
\tabletypesize{\footnotesize}
\tablecaption{Best-fit mass models}
\tablewidth{0pt}
\tablehead{\colhead{Model} & 
\colhead{$\sigma$ [${\rm km\,s^{-1}}$]} &
\colhead{$e$ or $\gamma$} &
\colhead{$\theta_e$ or $\theta_\gamma$ [deg]} & 
\colhead{$R_{\rm Ein}$ [arcsec]} & 
\colhead{$\mu_{\rm tot}$} & 
\colhead{$\Delta t$ [day]} }
\startdata
SIE         & $147$  & $0.235$ & $-47.1$ & $0.40$ & $6.0$ & $11.1$ \\
SIS+$\gamma$& $143$  & $0.096$ & $-47.2$ & $0.38$ & $6.1$ & $10.0$ \\
\enddata
\label{tab:model}
\end{deluxetable}

\clearpage

\begin{thebibliography}{}

\bibitem[Abazajian et al.(2003)]{abazajian03}
Abazajian, K., et al.\ 2003, \aj, 126, 2081

\bibitem[Abazajian et al.(2004)]{abazajian04}
Abazajian, K., et al.\ 2004, \aj, 128, 502

\bibitem[Abazajian et al.(2005)]{abazajian05}
Abazajian, K., et al.\ 2005, \aj, 129, 1755

\bibitem[Abazajian et al.(2009)]{abazajian09}
Abazajian, K.~N., et al.\ 2009, \apjs, 182, 543 

\bibitem[Adelman-McCarthy et al.(2006)]{adelman06}
Adelman-McCarthy, J.~K., et al.\ 2006, \apjs, 162, 38

\bibitem[Adelman-McCarthy et al.(2007)]{adelman07}
Adelman-McCarthy, J.~K., et al.\ 2007, \apjs, 172, 634

\bibitem[Adelman-McCarthy et al.(2008)]{adelman08} 
Adelman-McCarthy, J.~K., et al.\ 2008, \apjs, 175, 297 

\bibitem[Auger et al.(2008)]{auger08}
Auger, M.~W., Fassnacht, C.~D., Wong, K.~C., Thompson, D., 
Matthews, K., \& Soifer, B.~T.\ 2008, \apj, 673, 778 

\bibitem[Blackburne et al.(2008)]{blackburne08}
Blackburne, J.~A., Wisotzki, L., \& Schechter, P.~L.\ 2008, 
\aj, 135, 374 

\bibitem[Blanton et al.(2003)]{blanton03}
Blanton, M.~R., Lin, H., Lupton, R.~H., Maley, F.~M., 
Young, N., Zehavi, I., \& Loveday, J.\ 2003, \aj, 125, 2276 

\bibitem[Bolzonella et al.(2000)]{bolzonella03}
Bolzonella, M., Miralles, J.-M., \& Pell{\'o}, R.\ 2000, \aap, 363, 476 

\bibitem[Brammer et al.(2008)]{brammer08}
Brammer, G.~B., van Dokkum, P.~G., \& Coppi, P.\ 2008, \apj, 686, 1503 

\bibitem[Browne et al.(2003)]{browne03}
Browne, I.~W.~A., et al.\ 2003, \mnras, 341, 13 

\bibitem[Bruzual \& Charlot(1993)]{bruzual93}
Bruzual A., G., \& Charlot, S.\ 1993, \apj, 405, 538 

\bibitem[Castander et al.(2006)]{castander06}
Castander, F.~J., Treister, E., Maza, J., \& 
Gawiser, E.\ 2006, \apj, 652, 955 

\bibitem[Chae et al.(2002)]{chae02}
Chae, K.-H., et al.\ 2002, Physical Review Letters, 89, 151301 

\bibitem[Chiba et al.(2005)]{chiba05}
Chiba, M., Minezaki, T., Kashikawa, N., Kataza, H., \& 
Inoue, K.~T.\ 2005, \apj, 627, 53 

\bibitem[Coleman et al.(1980)]{coleman80}
Coleman, G.~D., Wu, C.-C., \& Weedman, D.~W.\ 1980, \apjs, 43, 393 

\bibitem[Doi et al.(2010)]{doi10}
Doi, M., et al.\ 2010, \aj, 139, 1628 

\bibitem[Fukugita \& Turner(1991)]{fukugita91}
Fukugita, M., \& Turner, E.~L.\ 1991, \mnras, 253, 99 

\bibitem[Fukugita et al.(1996)]{fukugita96}
Fukugita, M., Ichikawa, T., Gunn, J.~E., Doi, M., Shimasaku, K., 
\& Schneider, D.~P.\ 1996, \aj, 111, 1748 

\bibitem[Gunn et al.(1998)]{gunn98}
Gunn, J.~E., et al.\ 1998, \aj, 116, 3040 

\bibitem[Gunn et al.(2006)]{gunn06}
Gunn, J.~E., et al.\ 2006, \aj, 131, 2332 

\bibitem[Hayano et al.(2008)]{hayano08}
Hayano, Y., et al.\ 2008, \procspie, 7015, 25

\bibitem[Hayano et al.(2010)]{hayano10}
Hayano, Y., et al.\ 2010, \procspie, 7736, 21

\bibitem[Hogg et al.(2001)]{hogg01}
Hogg, D.~W., Finkbeiner, D.~P., Schlegel, D.~J., \& 
Gunn, J.~E.\ 2001, \aj, 122, 2129 

\bibitem[Ichikawa et al.(2006)]{ichikawa06}
Ichikawa, T., et al.\ 2006, \procspie, 6269, 38

\bibitem[Inada et al.(2008)]{inada08}
Inada, N., et al.\ 2008, \aj, 135, 496 

\bibitem[Inada et al.(2010)]{inada10}
Inada, N., et al.\ 2010, \aj, 140, 403 

\bibitem[Ivezi{\'c} et al.(2004)]{ivezic04}
Ivezi{\'c}, {\v Z}., et al.\ 2004, Astronomische Nachrichten, 325, 583 

\bibitem[Iye et al.(2004)]{iye04}
Iye, M., et al.\ 2004, PASJ, 56, 381  

\bibitem[Kashikawa et al.(2002)]{kashikawa02}
Kashikawa, N., et al.\ 2002, \pasj, 54, 819 

\bibitem[Kayo et al.(2010)]{kayo10}
Kayo, I., Inada, N., Oguri, M., Morokuma, T., 
Hall, P.~B., Kochanek, C.~S., 
\& Schneider, D.~P.\ 2010, \aj, 139, 1614 

\bibitem[Keeton et al.(1998)]{keeton98}
Keeton, C.~R., Kochanek, C.~S., \& Falco, E.~E.\ 1998, \apj, 509, 561 

\bibitem[Kobayashi et al.(2000)]{kobayashi00}
Kobayashi, N., et al. \ 2000, \procspie, 4008, 1056

\bibitem[Lagattuta et al.(2010)]{lagattuta10}
Lagattuta, D.~J., Auger, M.~W., \& Fassnacht, C.~D.\ 2010, \apjl, 716,
L185  

\bibitem[Leggett et al.(2006)]{leggett06}
Leggett, S.~K., et al.\ 2006, \mnras, 373, 781 

\bibitem[McKean et al.(2007)]{mckean07}
McKean, J.~P., et al.\ 2007, \mnras, 378, 109 

\bibitem[Minowa et al.(2005)]{minowa05}
Minowa, Y., et al.\ 2005, \apj, 629, 29 

\bibitem[Minowa et al.(2010)]{minowa10}
Minowa, Y., et al.\ 2010, \procspie, 7736, 122 

\bibitem[Oguri(2007)]{oguri07}
Oguri, M.\ 2007, \apj, 660, 1 

\bibitem[Oguri(2010)]{oguri10}
Oguri, M.\ 2010, \pasj, 62, 1017 

\bibitem[Oguri et al.(2006)]{oguri06}
Oguri, M., et al.\ 2006, \aj, 132, 999 

\bibitem[Oguri et al.(2008)]{oguri08}
Oguri, M., et al.\ 2008, \aj, 135, 512 

\bibitem[Padmanabhan et al.(2008)]{padmanabhan08}
Padmanabhan, N., et al.\ 2008, \apj, 674, 1217 

\bibitem[Peng et al.(2002)]{peng02}
Peng, C.~Y., Ho, L.~C., Impey, C.~D., \& Rix, H.-W.\ 2002, \aj, 124, 266 

\bibitem[Peng et al.(2006)]{peng06}
Peng, C.~Y., Impey, C.~D., Ho, L.~C., Barton, E.~J., 
\& Rix, H.-W.\ 2006, \apj, 640, 114 

\bibitem[Pier et al.(2003)]{pier03}
Pier, J.~R., Munn, J.~A., Hindsley, R.~B., Hennessy, G.~S., 
Kent, S.~M., Lupton, R.~H., 
\& Ivezi{\'c}, {\v Z}.\ 2003, \aj, 125, 1559 

\bibitem[Pindor et al.(2006)]{pindor06}
Pindor, B., et al.\ 2006, \aj, 131, 41 

\bibitem[Richards et al.(2002)]{richards02}
Richards, G.~T., et al.\ 2002, \aj, 123, 2945 

\bibitem[Reimers et al.(2002)]{reimers02}
Reimers, D., Hagen, H.-J., Baade, R., Lopez, S., \& Tytler, D.\ 2002,
\aap, 382, L26  

\bibitem[Rusin \& Kochanek(2005)]{rusin05}
Rusin, D., \& Kochanek, C.~S.\ 2005, \apj, 623, 666 

\bibitem[Sluse et al.(2008)]{sluse08}
Sluse, D., Courbin, F., Eigenbrod, A., \& Meylan, G.\ 2008, \aap, 492, L39 

\bibitem[Smith et al.(2002)]{smith02}
Smith, J.~A., et al.\ 2002, \aj, 123, 2121 

\bibitem[Stoughton et al.(2002)]{stoughton02}
Stoughton, C., et al.\ 2002, \aj, 123, 485

\bibitem[Suyu et al.(2010)]{suyu10}
Suyu, S.~H., Marshall, P.~J., Auger, M.~W., 
Hilbert, S., Blandford, R.~D., Koopmans, L.~V.~E., 
Fassnacht, C.~D., \& Treu, T.\ 2010, \apj, 711, 201 

\bibitem[Suzuki et al.(2008)]{suzuki08}
Suzuki, R., et al.\ 2008, \pasj, 60, 1347 

\bibitem[Treu \& Koopmans(2004)]{treu04}
Treu, T., \& Koopmans, L.~V.~E.\ 2004, \apj, 611, 739 

\bibitem[Tucker et al.(2006)]{tucker06}
Tucker, D.~L., et al.\ 2006, Astronomische Nachrichten, 327, 821 

\bibitem[Walsh et al.(1979)]{walsh79}
Walsh, D., Carswell, R.~F., \& Weymann, R.~J.\ 1979, \nat, 279, 381 

\bibitem[York et al.(2000)]{york00}
York, D.~G., et al.\ 2000, \aj, 120, 1579

\end{thebibliography}
\end{document}